\documentclass[prd,showpacs,superscriptaddress,twocolumn]{revtex4-1}
\usepackage{hyperref} \usepackage{graphicx} \usepackage{graphics}
\usepackage{amsmath} \usepackage{amssymb} \usepackage{amsthm}

\newcommand{\be}{\begin{equation}} \newcommand{\ee}{\end{equation}}
\newcommand{\bea}{\begin{eqnarray}} \newcommand{\eea}{\end{eqnarray}}
\begin{document}

\title{Nucleon Excited States in N$_f$=2 lattice QCD }

\author{C. Alexandrou} \affiliation{Department of Physics, University
  of Cyprus, P.O. Box 20537, 1678 Nicosia, Cyprus} \affiliation{The
  Cyprus Institute, P.O. Box 27456, 1645 Nicosia, Cyprus}
\author{T. Korzec} \affiliation{Institut f\"ur Physik, Humboldt
  Universit\"at zu Berlin, Newtonstrasse 15, 12489 Berlin, Germany}
\author{G. Koutsou} \affiliation{The Cyprus Institute, P.O. Box 27456,
  1645 Nicosia, Cyprus} \author{T. Leontiou} \affiliation{General
  Department, Frederick University, 1036 Nicosia, Cyprus}

\begin{abstract}
We investigate the excited states of the nucleon using $N_f=2$ twisted
mass gauge configurations with pion masses in the range of about
270~MeV to 450~MeV and one ensemble of $N_f=2$ Clover fermions at
almost physical pion mass. We use two different sets of variational
bases and study the resulting generalized eigenvalue problem.  We
present results for the two lowest positive and negative parity
states.
\end{abstract}

\maketitle

\section{Introduction}
Understanding the excitation spectrum of hadrons, including that of
the proton is still a challenge. In particular, the $P_{11}(1440{\rm
  MeV})$ positive parity resonance known as the Roper, still remains a
puzzle having a mass lower than the negative parity state
$S_{11}(1535{\rm MeV})$. This ordering is contrary to the prediction
of constituent quark models where the negative parity state is lower
in mass than $P_{11}$.  Lattice QCD simulations have recently
reproduced the mass of the low-lying baryon states using gauge
configurations with pions having mass close to the physical
value~\cite{Durr:2008zz,Alexandrou:2008tn}. In these studies volume
and cut-off effects have been taken into account by performing the
calculation at different volumes and lattice spacings.  Contrary to
the low-lying baryon states the study of excited states has not yet
reached the same level of maturity.  In order to extract excited state
energies, a robust analysis of simulation data keeping systematic
errors under control is needed.

The study of excited states is mostly based on the variational
principle, which was first applied to extract glueball
masses~\cite{Michael:1982gb}. One considers a number of interpolating
fields as a variational basis and a generalized eigenvalue problem
(GEVP) is defined, which yields the low-lying energy levels. The GEVP
has been applied recently to study hadron spectroscopy by a number of
lattice
groups~\cite{Basak:2007kj,Gattringer:2008be,Gattringer:2008vj,Mahbub:2010jz,Bulava:2010yg}.
A crucial question of such an approach is the convergence of the
energy levels to the true value. This was first addressed in a paper
by L\"uscher and Wolff~\cite{Luscher:1990ck} and recently by the
ALPHA-collaboration~\cite{Blossier:2009kd}.  In this work, we explore
the variational approach as put forward by the ALPHA-collaboration to
study the excited states of the nucleon in the positive and negative
parity channels. We examine two types of nucleon interpolating fields
as well as different levels of Gaussian smearings.  The approach
proposed by the ALPHA-collaboration is compared with the standard
GEVP, where the reference time $t_0$ is kept fixed at a small value.
The main outcome of this comparison is that, within the current
statistical accuracy typically used for baryon calculations, namely
${\cal O}(10^2)$ configurations, we do not see any improvements to the
standard analysis.  Having established at one ensemble of twisted mass
fermions that the standard generalized eigenvalue approach performs
equally well, we adopt it for the other ensembles. In the positive
parity channel we include in the variational basis interpolating
fields with a large and small number of iterations in the Gaussian
smearing. This is crucial to reproduce a state with lower energy
closer to the Roper state. As argued in
Refs.~\cite{Mahbub:2010jz,Mahbub:2010rm,Roberts:2012em} a linear
combination of interpolating fields corresponding to a small and large
root mean square radius (rms) produces a wavefunction with a node
having potentially a larger overlap with the Roper state. We indeed
observe a lowering in the energy of the first excited state when
including an interpolating field with a large rms radius.

 We analyze a total of five ensembles of $N_f=2$ twisted mass fermions
 with pion mass in the range of about 270~MeV to 450~MeV and lattice
 spacing $a=0.089$~fm determined from the nucleon
 mass~\cite{Alexandrou:2008tn}. Cut-off effects on the mass of the
 nucleon and hyperons were examined in
 Refs.~\cite{Alexandrou:2008tn,Alexandrou:2009qu} respectively using,
 in addition to the one used here, two smaller lattice spacings. The
 conclusion was that cut-off effects were within the statistical
 errors and one could take the continuum limit assuming negligible
 ${\cal O}(a^2)$ effects. Therefore, in this work, we limit ourselves
 to studying only one lattice spacing.  In addition, we analyze an
 ensemble of $N_f=2$ Clover fermions with pion mass $m_\pi\sim
 160$~MeV and lattice spacing $a\simeq0.073$~fm.

The paper is organized as follows: In section~\ref{sec:sim_details} we
give the details of the simulations, in section~\ref{sec:gevp} we
compare results using different variational bases and analysis
approaches using an ensemble of twisted mass fermions with $m_\pi\sim
300$~MeV, in section~\ref{sec:results} we give our results and in
section~\ref{sec:conclusions} we summarize our findings and give our
conclusions.

\section{Simulation Details}
\label{sec:sim_details}
The input parameters of the calculation using $N_f=2$ twisted mass
fermions, namely $\beta$, $L/a$ and $a\mu$ are summarized in
Table~\ref{Table:params}. These are the same configurations already
used in the analysis of the low-lying baryon spectrum
\cite{Alexandrou:2009qu}, where more details regarding the twisted
mass formulation can be found. The corresponding lattice spacing $a$
and the pion mass values, spanning a mass range from 270~MeV to
450~MeV, are taken from Ref.~\cite{Alexandrou:2008tn}. We note that
for baryon masses we use the lattice spacing determined from the
nucleon mass, which is consistent with the one extracted from
$f_\pi$~\cite{Urbach:2007rt}.
\begin{table}[h]
\begin{center}
\begin{tabular}{c|lllll}
\hline\hline \multicolumn{5}{c}{$\beta=3.9$, $a=0.089(1)(5)$~fm from
  the nucleon mass}\\ \multicolumn{5}{c}{ ${r_0/a}=5.22(2)$}\\\hline
$24^3\times 48$, & $a\mu$ & 0.0040 & 0.0064 & 0.0085 \\ $L=2.05$~fm &
No. of confs & 400 & 400 & 348 \\ & $m_{\pi^{\pm}}$~(GeV) & 0.3131(16)
& 0.3903(9) & 0.4470(12) \\ & $L\, m_{\pi^{\pm}}$ & 3.25 & 4.05 & 4.63
\\ $32^3\times 64$, & $a\mu$ & 0.003 & 0.004 & & \\ $L=2.74$~fm &
No. of confs & 400 & 250 & & \\ & $m_{\pi^{\pm}}$~(GeV) & 0.2696(9) &
0.3082(6) & & \\ & $L\,m_{\pi^{\pm}}$ & 3.74 & 4.28 & & \\\hline
\hline
\end{tabular}
\caption{Input parameters ($\beta,L,\mu$) of our lattice calculation
  and corresponding lattice spacing ($a$), pion mass ($m_{\pi}$) and
  number of gauge field configurations used. The values of the pion
  mass in physical units were obtained using the lattice spacing
  determined from $f_\pi$, namely $a=0.0855(6)$~fm.}
\label{Table:params}
\end{center}
\vspace*{-.0cm}
\end{table} 

\begin{figure}
  \begin{center}
    \includegraphics[width=1\linewidth]{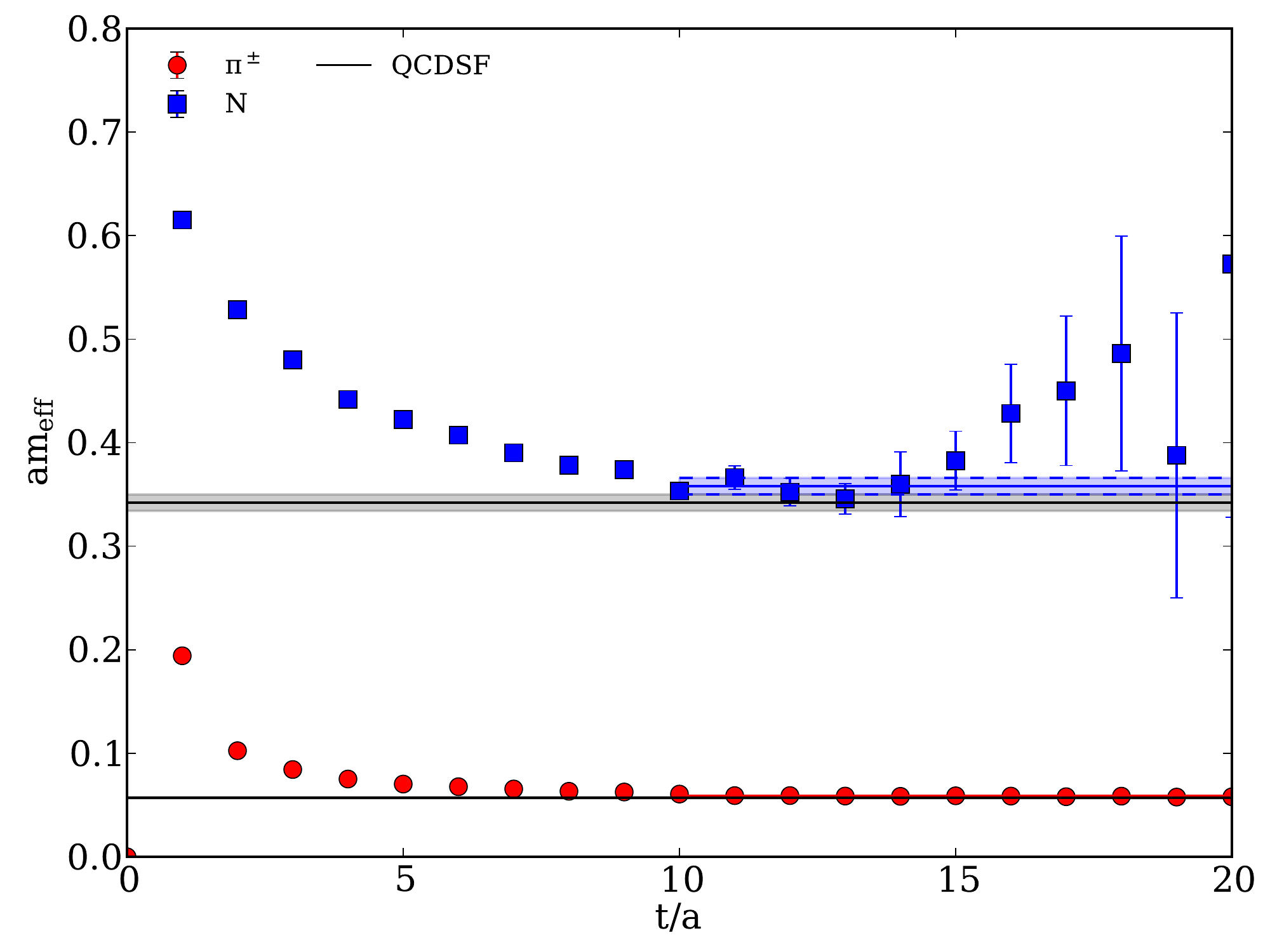}
    \caption{The pion (red circles) and nucleon (blue squares)
      effective masses in the non-unitary setup as described in the
      text, compared to their values in the unitary theory (solid
      black line) computed by QCDSF~\cite{Bali:2012qs}. The value of
      $\kappa$ in the non-unitary setup was tuned to reproduce the
      pion mass in the unitary theory.}
    \label{fig:meff_tuned}
  \end{center}
\end{figure}

Apart from the twisted mass fermion ensembles given in
Table~\ref{Table:params} we also analyze an ensemble of $N_f$ = 2
Clover fermion configurations produced by the QCDSF collaboration. We
use the $48^3\times64$ ensemble with near-physical pion mass of
$m_{\pi}\simeq160$~MeV, at $\beta=5.29$ for which the lattice spacing
has been determined to be
$a=0.0728(5)(19)$~fm~\cite{Bali:2012qs}. This yields a value for
$L\,m_\pi\simeq2.8$. We smear the links that enter the Dirac operator
with three iterations of APE smearing~\cite{Albanese:1987ds} to reduce
gauge noise and set the clover term to its tree-level value
i.e. $c_{SW}=1$. Smearing the links in this way changes $\kappa_{\rm
  crit}$. We therefore tune the value of the hopping parameter
$\kappa$ as described in~\cite{Durr:2012dw} to match the pion mass in
the unitary theory. A comparison of the pion and nucleon effective
masses, $am_{\rm eff}(t)\equiv C(t)/C(t+1)$, in the unitary theory and
after tuning is shown in Fig.~\ref{fig:meff_tuned}. As can be seen,
the mass of the nucleon in the non-unitary theory agrees with the one
obtained in the unitary theory. Note that one has to allow 10 time
slices or about 0.7~fm to ensure that excited states have been
sufficiently suppressed.  This is a rather large time interval given
that the mass gap between the ground and the excited state estimated
from a double exponential fit, yields a suppression factor of ${\cal
  O}(e^{-4})$, which means that there is a substantial overlap of the
standard nucleon interpolating field with higher excited states.
\section{The variational Method}
\label{sec:gevp}
 The standard extraction of the ground state energy from the large
 time limit of Euclidean two-point correlation functions relies on the
 fact that they are expressed as a sum of the energy eigenstates of
 QCD that exponentially decay as a function of the time with a rate
 proportional to the energy. The variational method provides an
 approach for extracting, besides the lowest energy state, the
 low-lying excited states from Euclidean correlation functions. A
 variational basis is constructed by using different interpolating
 fields $\chi$ with the quantum numbers of the particular state of
 interest, which in this work is the nucleon.  Applying the
 variational principle one can determine the superposition of states
 that correspond to the low-lying nucleon states. One variational
 basis is obtained by considering two different spin combinations of
 nucleon interpolating fields, namely \bea \chi_1=(u^{\rm
   T}C\gamma_5d)u\,\,\, {\rm and}\,\,\,\chi_2=(u^{\rm
   T}Cd)\gamma_5u. \label{corrs} \eea The nucleon interpolator,
 $\chi_1$, is well known to have a good overlap with the ground state
 of the nucleon, while the $\chi_2$ interpolator vanishes in the
 non-relativistic limit and thus has a small overlap with the nucleon
 ground state, which is a motivation to include it in a variational
 basis to study the excited states.  In addition, the variational
 basis is enlarged by considering different Gaussian smearings using
 similar parameters to those used in Ref.~\cite{Blossier:2009kd}, as
 well as an interpolating field with larger smearing, which maybe
 needed for isolating the Roper.  The correlation matrix considered
 here, thus, has the general form:
\begin{align}
\label{gevp1}
C^{\pm}_{a_ib_j}(t)&=\sum_{\bf
  x}\textrm{Tr}[\frac{1}{4}(1\pm\gamma_0)\langle\chi_a^{(i)}({\bf
    x},t)\bar{\chi}_b^{(j)}({\bf 0},0)\rangle]
\nonumber\\ &=\sum_{n=0}^\infty e^{-E_nt
}\textrm{Tr}[\frac{1}{4}(1\pm\gamma_0)\langle 0|\chi_a^{(i)}|n\rangle
  \langle n|\chi_b^{(j)}|0\rangle]\,,&\nonumber\\ &\begin{array}{l}
  i,j=1,\ldots, N\\a,b=1,2,
\end{array}
\end{align}
where the trace is taken over Dirac indices and $C^+(t)$ ($C^-(t)$)
yields the positive (negative) parity
correlator~\cite{Lee:1998cx}. The states $|n\rangle$ are eigenstates
of the Hamiltonian with $E_n< E_{n+1}$ and we have assumed that the
temporal extent of the lattice is large enough to neglect
contributions due to the finite size of the temporal direction. The
indices $i$ and $j$ on the correlation matrix $C^{\pm}(t)$ correspond
to different levels of Gaussian smearing and $a$ and $b$ to $\chi_1$
and $\chi_2$.

\subsection{Variational basis with different gaussian smearing levels of $\chi_1$ }
 In this subsection, we perform an analysis using as a variational
 basis $\chi_1$ with a number of different smearing levels.  The
 variational basis is constructed using $N$ different Gaussian
 smearing levels of this interpolating field. The GEVP is defined by
 the generalized eigen-equation
\begin{align}
C(t)v_n(t,t_0)&=\lambda_n(t,t_0)C(t_0)v_n(t,t_0),\nonumber\\ n&=1,\ldots,
N, \, t>t_0\,,
\label{GEVP}
\end{align}
where $E_n=\lim_{t\rightarrow \infty} -\partial_t \log
\lambda_n(t,t_0)$. The corrections to $E_n$ decrease exponentially
like $e^{-\Delta E_n t}$ where $\Delta E_n=\min_{m\neq
  n}|E_m-E_n|$~\cite{Luscher:1990ck} for fixed $t_0$. In
Ref.~\cite{Luscher:1990ck,Blossier:2009kd} it was shown that if one
varies $t_0$ such that $t_0\ge t/2$ then the correction is ${\cal
  O}(e^{-\Delta E_{{N},n} t})$ with $\Delta E_{m,n}=E_m-E_n$ ensuring
a greater rate of convergence. In this section, we examine the benefit
of this relation for extracting the low-lying states in the nucleon
sector. A related work exploring the dependence of the GEVP on the
reference time is also examined in Ref.~\cite{Menadue:2013kfi} where
recurrence relations are obtained. The variational method has also
been extensively used to study the excited nucleon spectrum by the
Berlin-Graz-Regensburg (BGR) collaboration~\cite{Engel:2013ig}.

 We apply Gaussian smearing to each quark field, $q({\bf
   x},t)$~\cite{Gusken:1989,Alexandrou:1992ti}, entering $\chi_1$. The
 smeared quark field is given by $q^{\rm smear}({\bf x},t) = \sum_{\bf
   y} F({\bf x},{\bf y};U(t)) q({\bf y},t)$ using the gauge invariant
 smearing function \be F({\bf x},{\bf y};U(t)) = (1+\alpha
 H)^{n_s}({\bf x},{\bf y};U(t)), \ee constructed from the hopping
 matrix understood as a matrix in coordinate and color space \be
 H({\bf x},{\bf y};U(t))= \sum_{i=1}^3 \biggl( U_i({\bf
   x},t)\delta_{{\bf x,y}-a\hat \imath} + U_i^\dagger({\bf x}-a\hat
 \imath,t)\delta_{{\bf x,y}+a\hat \imath}\biggr).  \ee
 
Smearing is applied at the fermion source and sink. Following
Ref.~\cite{Blossier:2009kd} we consider values of the smearing
parameters $\alpha=0.1$ and $n_s=$0, 22, 45, 67 and 135.  These
smearing parameters produce a source with a root mean square radius in
lattice units of 0, 1.96, 2.72, 3.25 and 4.48, respectively. These
different smearing levels are labeled by the superscript $i=1,\ldots,
5$ on $\chi^{(i)}$. We will refer to this basis as {\it basis A}. The
resulting correlation matrices are symmetrized. We use 150 twisted
mass configurations with $\beta$=3.9, a$\mu$ = 0.004 or m$_\pi\sim
308$~MeV on a 32$^3$ $\times$ 64 lattice. In addition, we also
construct a $3\times3$ GEVP with a variational basis that includes a
heavily smeared interpolating field. For the latter basis, referred to
as {\it basis B}, the values of the smearing parameters are
$\alpha=4.0$ and $n_s=$10, 50, 180 producing a source with rms radius
in lattice units of 2.36, 4.87 and 8.60.  We analyze 200
configurations of the same ensemble for this variational basis. These
smearing levels will be labeled by the superscript $i = 6,7$ and 8 on
$\chi^{(i)}$. Although the rms for $i=$6 and $i=$7 is similar to $i=$1
and $i=$5 this new set contains the heavily smeared basis, $i$=8.

\begin{figure}[h!]
  \centering \includegraphics[width=\linewidth]{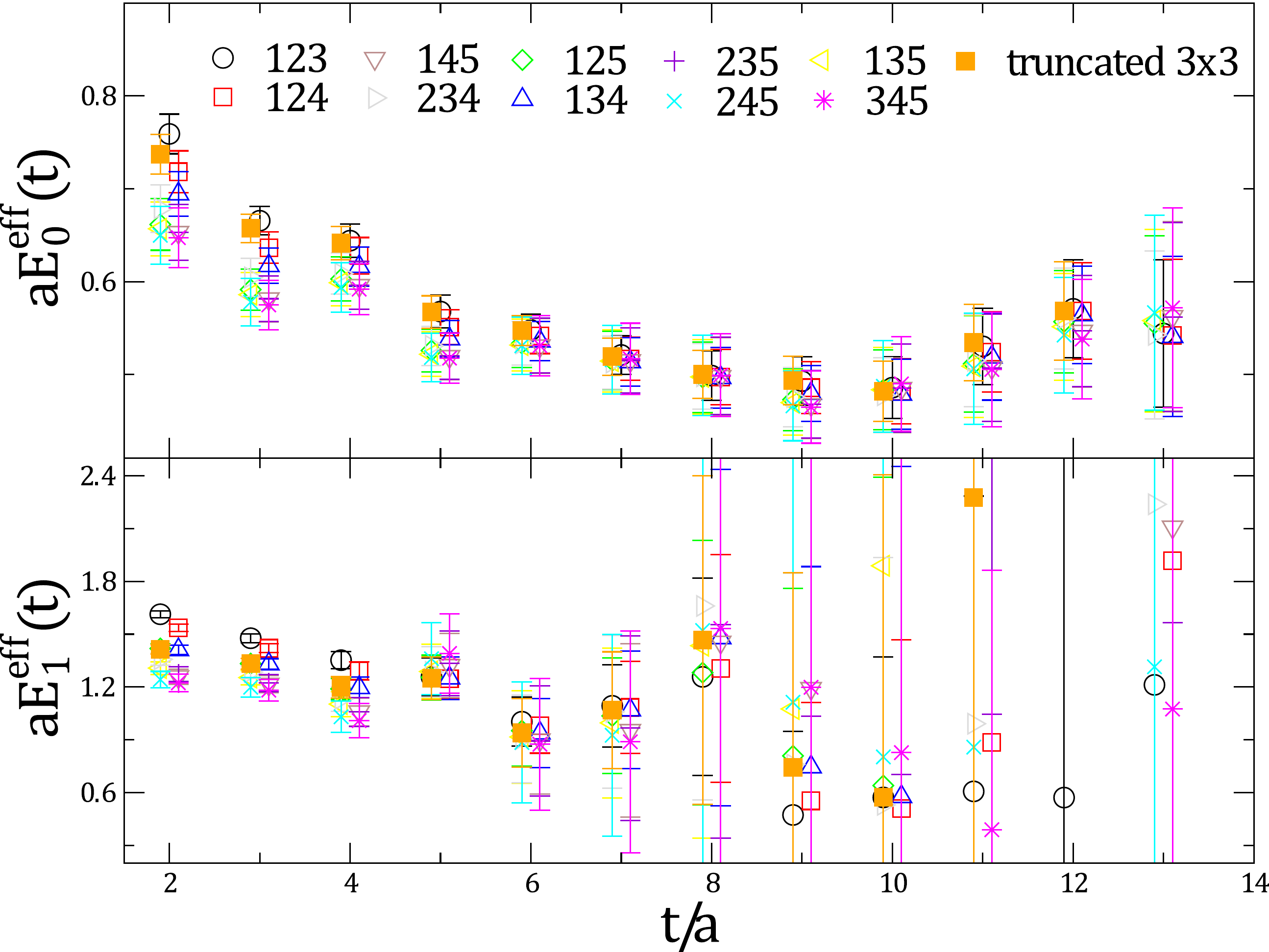}
  \caption{\label{best_basis} The effective mass for the ground
    ($E_0$) and first excited ($E_1$) states resulting from a
    $3\times3$ GEVP using {\it basis A}.  A $3\times3$ correlation
    matrix was constructed out of different interpolating fields
    $\chi_1^{(i)}$ by applying a different number of Gaussian smearing
    iterations on $\chi_1$.  The numbers in the legend give the
    combination of the three values of $n_s$ used to construct the
    basis.  The effective energy levels resulting from a truncated
    $3\times3$ GEVP constructed using Eq.~(\ref{truncated}) are also
    shown.  This analysis was carried out using 150 configurations of
    twisted mass fermions at $\beta$=3.9, a$\mu$ = 0.004
    (m$_\pi\sim308$~MeV) on a 32$^3$ $\times$ 64 lattice. }
\end{figure}

\begin{figure}[h!]
  \centering \includegraphics[width=\linewidth]{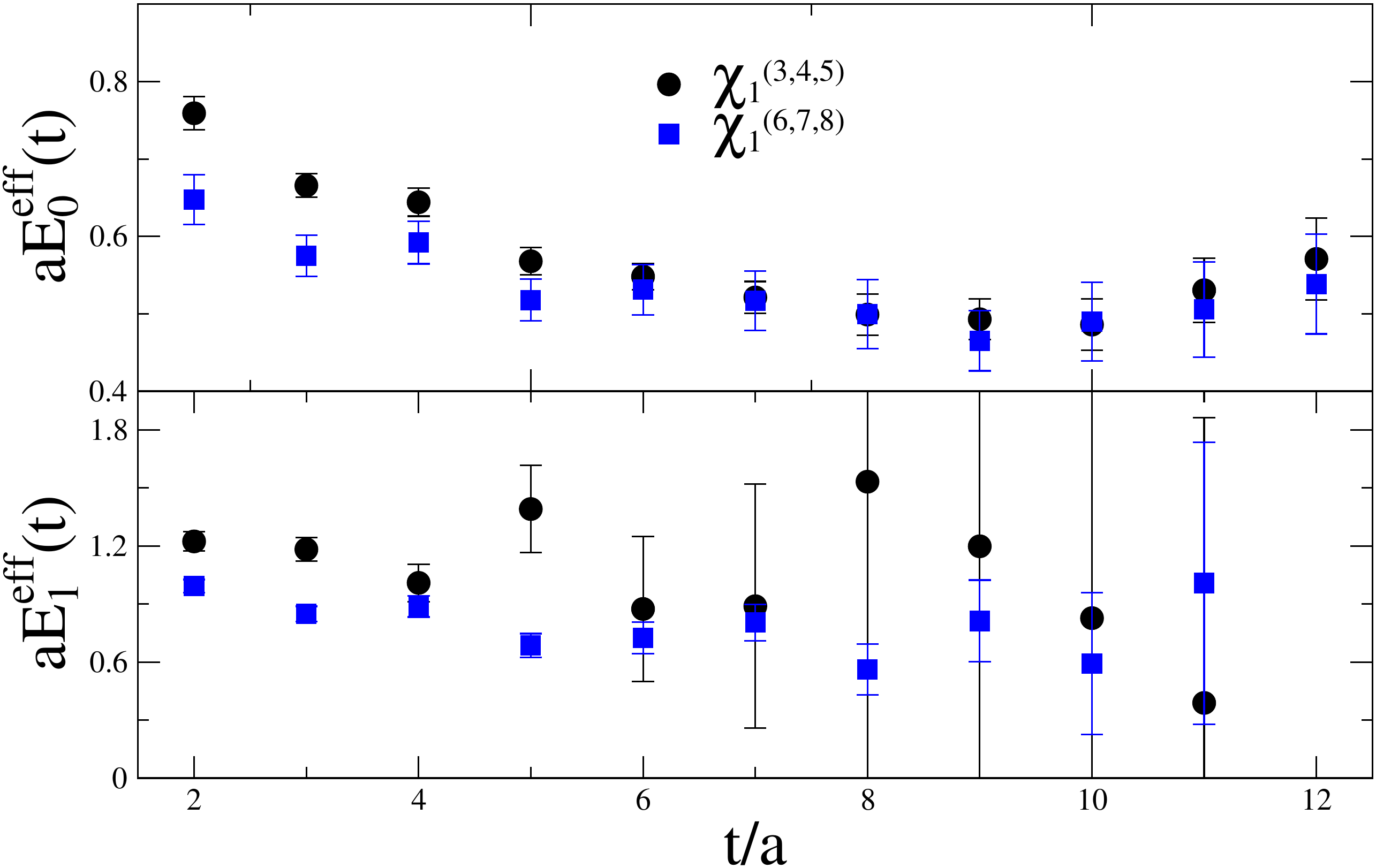}
  \caption{\label{best_basis2} The effective mass for the ground
    ($E_0$) and first excited ($E_1$) states resulting from a
    $3\times3$ GEVP using {\it basis A} (interpolating fields
    $\chi_1^{(3)}$, $\chi_1^{(4)}$ and $\chi_1^{(5)}$ ) using 150
    gauge configurations (black filled circles) and {\it basis B}
    using 200 gauge configurations (blue filled squares) of twisted
    mass fermions at $\beta$=3.9, a$\mu$ = 0.004 (m$_\pi\sim308$~MeV)
    on a 32$^3$ $\times$ 64 lattice. }
\end{figure}

\begin{figure}[!h]
  \centering \includegraphics[width=\linewidth,
    keepaspectratio]{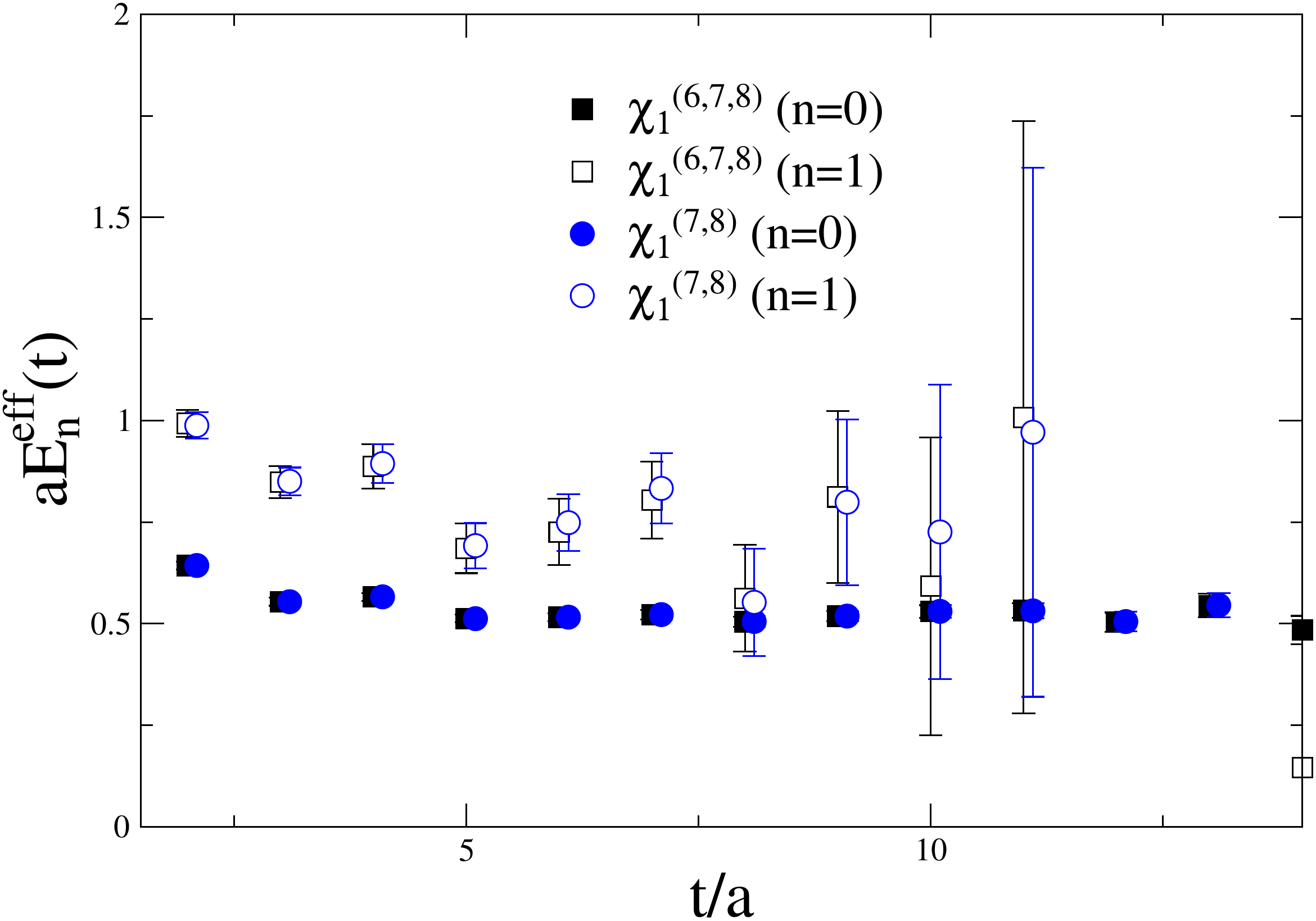}
  \caption{The effective mass for the ground and first excited states
    from the best choice of $2\times2$ and $3\times3$ GEVPs
    corresponding to the highest level of Gaussian smearing i.e. from
    $\{\chi_1^{(7)},\chi_1^{(8)}\}$ for the 2$\times$2 GEVP and
    $\{\chi_1^{(6)},\chi_1^{(7)},\chi_1^{(8)}\}$ for the 3$\times 3$
    GEVP. The test was carried out using 200 configurations of the
    twisted mass ensemble with $\beta$=3.9, a$\mu$ = 0.004
    (m$_\pi\sim308$~MeV) on a 32$^3$ $\times$ 64 lattice
   \label{best_plateaus}}
\end{figure}

Let us first examine the role of $t_0$ and the advantage of using
these different smearing levels.  We consider several different
correlation matrices of the positive parity correlator $C^+_{1_i\>
  1_j}(t)$ constructed from $\chi_1^{(i)}$ for different smearing
levels $i=1,\ldots,5$ in order to examine both the role of varying
$n_s$ and/or the dimensionality of the GEVP. In Fig.~\ref{best_basis}
we show the effective mass for the ground and first excited states
resulting from a GEVP analysis of all possible 3$\times$3 correlation
matrices fixing $t_0/a=1$. We are looking for the combination of
interpolating fields that gives the fastest convergence to the
two-lowest levels $E_0$ and $E_1$ i.e.  to the earliest onset of a
plateau behavior. From this analysis it is evident that using the
higher smearing levels improves convergence allowing us to fit to a
constant starting from time-slice $t/a=5$ for the ground state and
from time-slice $t/a=4$ for the first excited state. The condition
number of this 3$\times$3 GEVP ranges from $10^4$ (when
$\chi_1^{(1)}$, $\chi_1^{(2)}$ and $\chi_1^{(3)}$ are used) up to
$10^6$ (when $\chi_1^{(1)}$, $\chi_1^{(3)}$ and $\chi_1^{(5)}$ are
used).

Next we examine the role of increasing the level of smearing and
compare the results obtained from the above analysis with a $3\times
3$ GEVP using {\it basis B}. In Fig.~\ref{best_basis2} we show the
effective mass for the ground and first excited states resulting from
a 3$\times$3 GEVP for both {\it basis A} and {\it basis B}. Using {\it
  basis B} we observe faster convergence to ground state and a
lowering in the value of the excited state mass.  The condition number
for {\it basis B} is in the order of $10^6$. Furthermore, increasing
the level of smearing beyond $n_s=$180 does not result in any further
lowering of the energy of the excited state but only leads to larger
statistical errors. In fact the condition number of the correlation
matrix gets worse increasing rapidly to ${\cal O}(10^9)$ when we use
$n_s=300$.  The comparison of these results indicates that for the
study of the positive parity states {\it basis B} is more suitable
than {\it basis A}.

Apart from making a choice of the appropriate basis by trying
different combinations of Gaussian smearing we also try a truncation
scheme where the $5\times5$ correlation matrix is projected to an
$m\times m$ matrix, $C^{m\times m}(t)$, with $m<N$ by using the $m<5$
eigenvectors belonging to the $m$ largest eigenvalues of $C(t_0)$ as
follows
\begin{align}
C^{N\times N}(t_0)b=\Lambda b, \hspace*{0.5cm} C_{kj}^{m\times
  m}(t)=b^\dagger_{ki_1}C^{N\times
  N}_{i_1i_2}(t)b_{i_2j},\nonumber\\k,j=1, \ldots,m,\hspace*{0.5cm}
i_1,i_2=1, \ldots,N ,
\label{truncated}
\end{align}
where $\Lambda_{jk}=\delta_{jk} e^{-E_j t_0}$ is an $N\times N$ matrix
with the eigenvalues of $C^{N\times N}(t_0)$ as its diagonal elements
and $b$ an $N\times N$ matrix with the corresponding eigenvectors. We
additionally tried this truncation scheme with various values of
$t_0/a$, namely $t_0/a$ = 1,...,4 and the results obtained are found
to be statistically equivalent. The resulting effective masses
extracted from the truncated $3\times 3$ matrix using {\it basis A}
are included in Fig.~\ref{best_basis} and do not show any improved
convergence.

The effect of reducing the dimension of the GEVP to $2\times2$ can be
seen in Fig. \ref{best_plateaus}. The quality of the plateaus for the
first two states is not affected as compared to those extracted using
the $3\times 3$ correlation matrix with $\chi_1^{(6)}$, $\chi_1^{(7)}$
and $\chi_1^{(8)}$.

\begin{figure}[h!]
  \includegraphics[width=\linewidth]{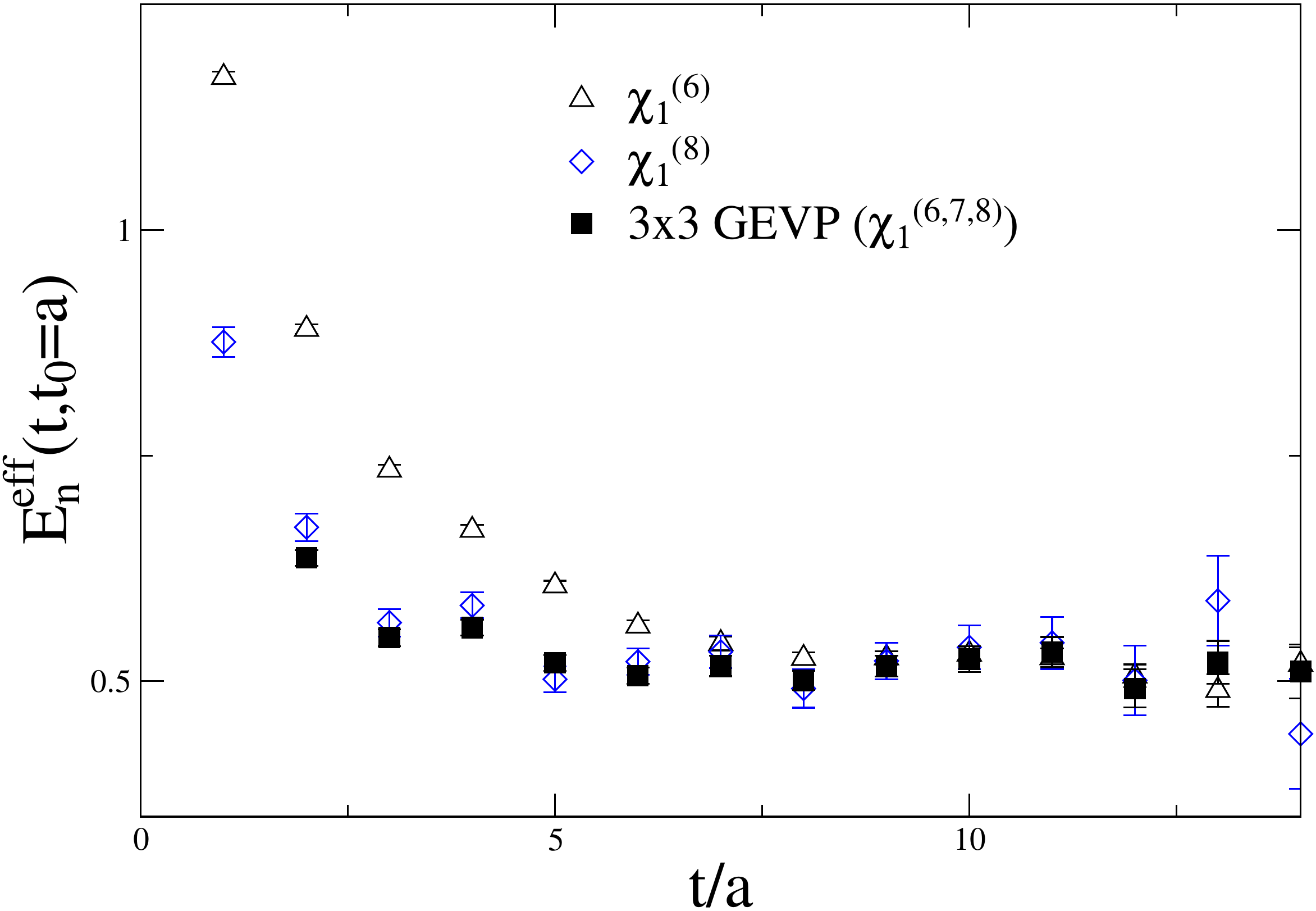}
\caption{\label{ground} The effective mass for the ground state for
  $t_0/a=1$. Results shown are extracted from the GEVP with {\it basis
    B} and from the correlators $C_{1_6\> 1_6}$ and $C_{1_8\> 1_8}$.}
\end{figure}

In Fig.~\ref{ground} we compare the results obtained using the GEVP
analysis to those extracted using a single interpolating field
$\chi_1^{(i)}$, i.e. the trivial $1\times 1$ GEVP. For the ground
state, using just the $\chi_1^{(8)}$ interpolating field yields the
same quality plateau as that obtained from the $3\times 3$ correlation
matrix analysis within {\it basis B}.

For the two lowest states in the positive channel we also study the
resulting eigenvectors in order to understand/verify the mixture of
the various $\chi_1^{(i)}$ contributing in the optimized interpolating
field for each state. Identifying the optimum combination of
$\chi_1^{(i)}$ extracted from the GEVP analysis is useful if one wants
to calculate the matrix elements of any operator using the optimal
interpolating field that best suppresses the contribution of excited
states.

\begin{figure}
  \centering \includegraphics[width=\linewidth]{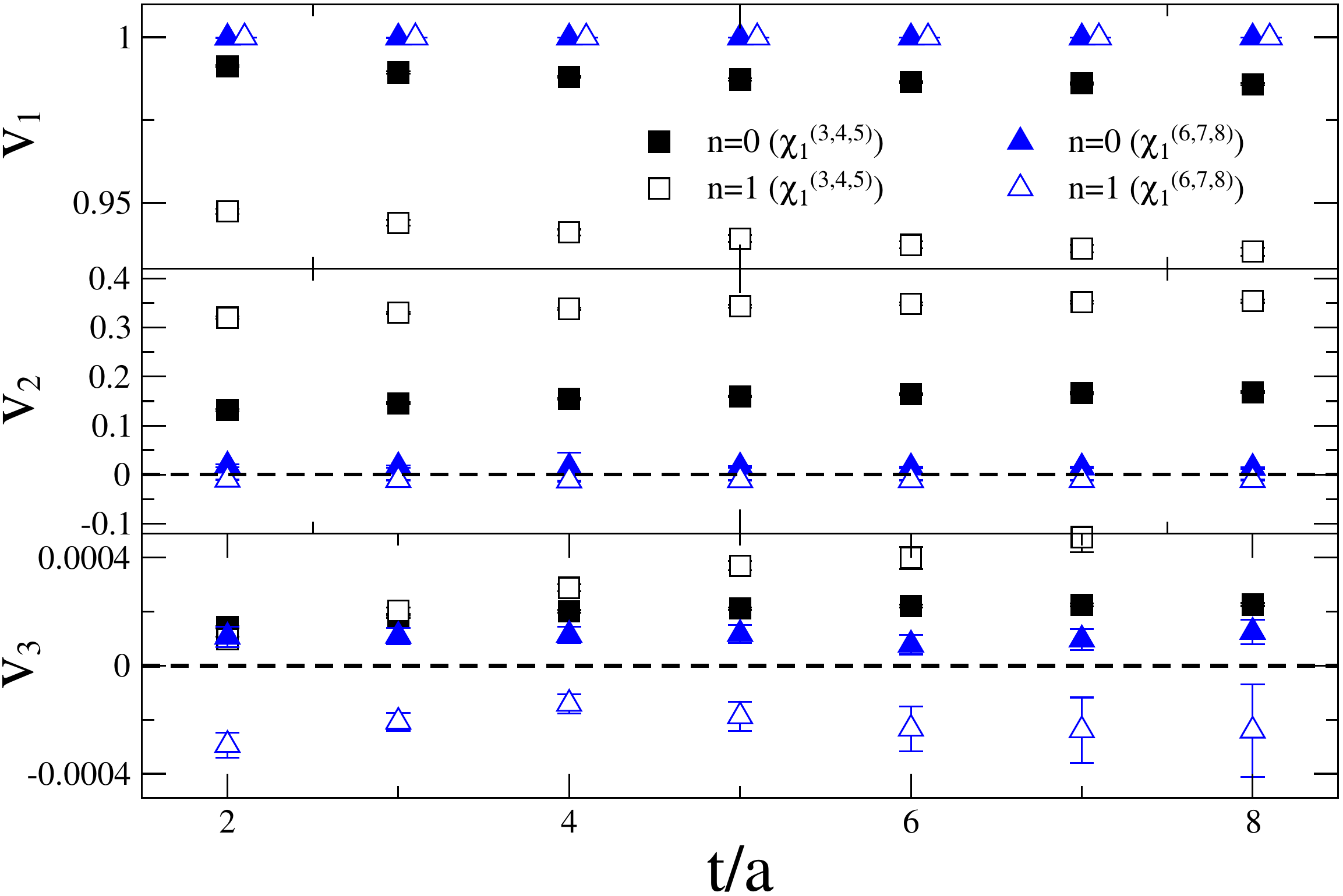}
  \caption{\label{vectors} The components of the eigenvector for the
    ground and first excited states at $t_0/a=1$. The results are
    extracted from GEVP analyses of the 3$\times$3 correlation
    matrices $C_{1_i1_j}$, $i,j=3,4,5$ and $C_{1_i1_j}$, $i,j=6,7,8$
    ({\it basis B}). }
\end{figure}

In Fig.~\ref{vectors} we show the three components $V_1, V_2$ and
$V_3$ of the eigenvector for the ground and excited state in the
positive parity channel determined from the $3\times3$ correlation
matrix for {\it basis A} (interpolating fields $\chi_1^{(3)}$,
$\chi_1^{(4)}$ and $\chi_1^{(5)}$) and for {\it basis B}
(interpolating fields $\chi_1^{(6)}$, $\chi_1^{(7)}$ and
$\chi_1^{(8)}$).  The interpolating field with the maximum overlap
with the ground state is given by $\chi_{\rm eff} =
\tilde{v}_1\chi_1^{(5)}+\tilde{v}_2\chi_1^{(4)}+\tilde{v}_3\chi_1^{(3)}$,
or equivalently by $\chi_{\rm eff} =
\tilde{v}_1\chi_1^{(8)}+\tilde{v}_2\chi_1^{(7)}+\tilde{v}_3\chi_1^{(6)}$,
where $\tilde{v}$ is the large-time limit of $V$ i.e.
$\tilde{v}(t_0)=\lim_{t\rightarrow \infty} V(t,t_0)$.  It is evident
that in the case of {\it basis B} one of the eigenvector component
enters in with the opposite sign from the other two thus providing the
possibility for a nodal structure, not possible with {\it basis A}.
Opposite signs for the eigenvectors are also obtained if we analyze a
$2\times2$ correlation matrix, as long as interpolator $\chi_1^{(8)}$
is used together with either $\chi_1^{(6)}$ or $\chi_1^{(7)}$.

\begin{figure}[h!]
  \centering
  \includegraphics[width=\linewidth]{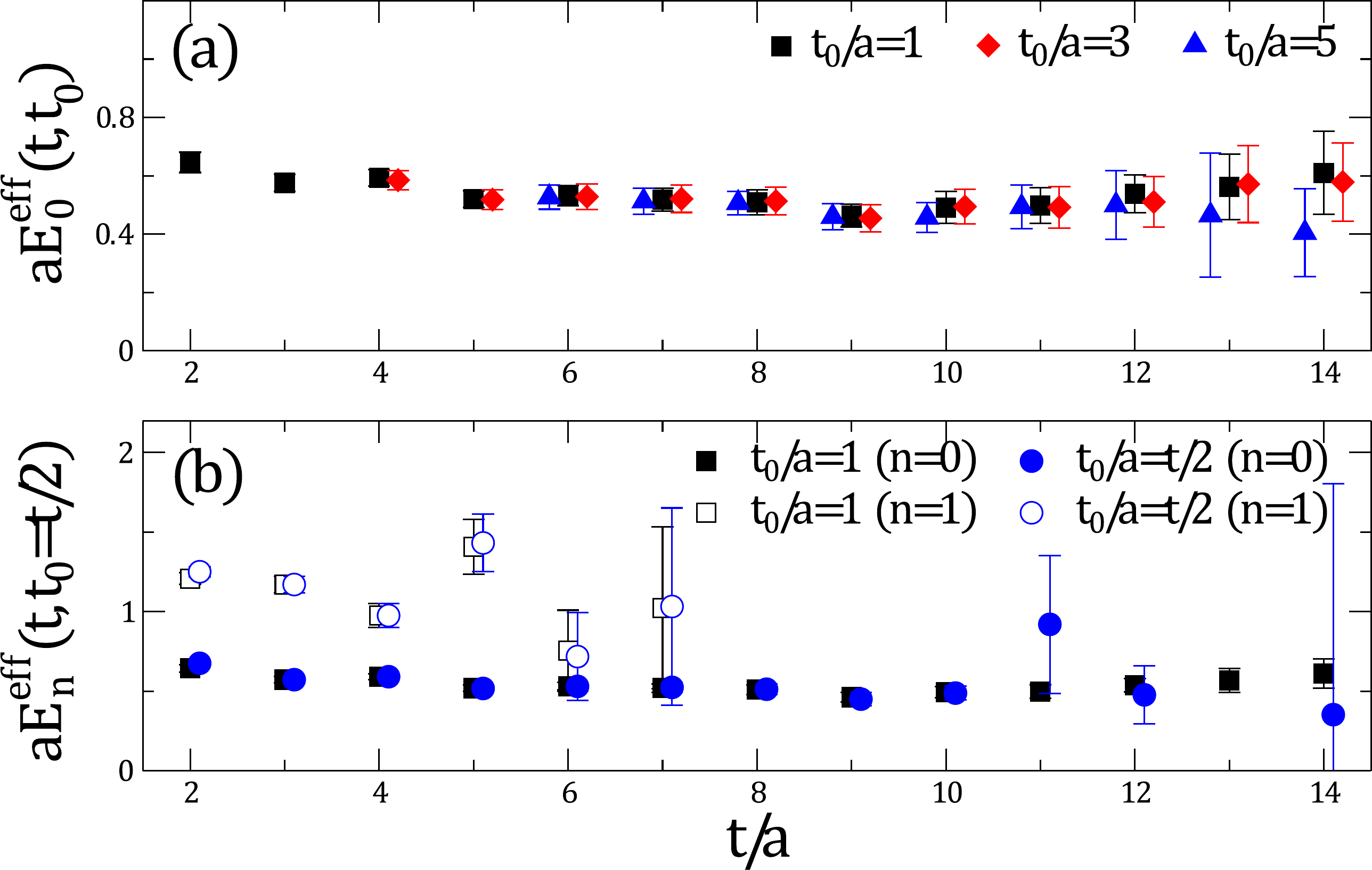}
  \caption{\label{t0} (a) The effective mass for the ground state for
    various choices of $t_0$. Results are shown for the 3$\times$ 3
    GEVP with the most smeared interpolating fields within {\it basis
      A}. (b) The effective mass for the ground and first excited
    states with a fixed value for $t_0$ (squares) and with the
    condition $t_0=t/2$ (circles) for the ground (filled symbols) and
    first excited state (open symbols).  Values have been slightly
    shifted in time in order to aid the comparison.}
\end{figure}

Let us next vary $t_0$ as suggested in Ref.~\cite{Blossier:2009kd},
shown to lead to an improvement in the determination of the ground
state by successfully suppressing excited state contamination for
certain mesonic systems. In Fig.~\ref{t0} we show results obtained at
fixed $t_0/a=1$ as well as results obtained by varying $t_0$ using
{\it basis A}. Within the statistical accuracy of our analysis, we see
consistent results for the three values of $t_0/a=1,\> 3$, and 5
considered. Furthermore, we allow $t_0$ to vary for every value of $t$
and in particular we apply the condition $t_0\ge t/2$ as suggested in
Ref.~\cite{Blossier:2009kd}. We show results for the ground and first
excited states in the positive parity channel for the case $t_0=t/2$,
where we observe no change in the plateau range within the present
statistics.  For these nucleon states and within the present accuracy,
this analysis does not show an improvement, a result that is also
valid for the variational {\it basis B}.  Our conclusion is that for
the low-lying nucleon spectrum, where the energy gap is not
particularly small, the variation of $t_0$ that has been shown in
Ref.~\cite{Blossier:2009kd} to reduce the systematic error is not
observed here at least within the limitation of our statistics.
Keeping $t_0\ge t/2$ comes at the cost of increased statistical
uncertainty. In our case, this increase is large and we find that the
highest overall precision is obtained by keeping $t_0/a=1$.

\begin{figure}[h!]
  \centering \includegraphics[width=\linewidth]{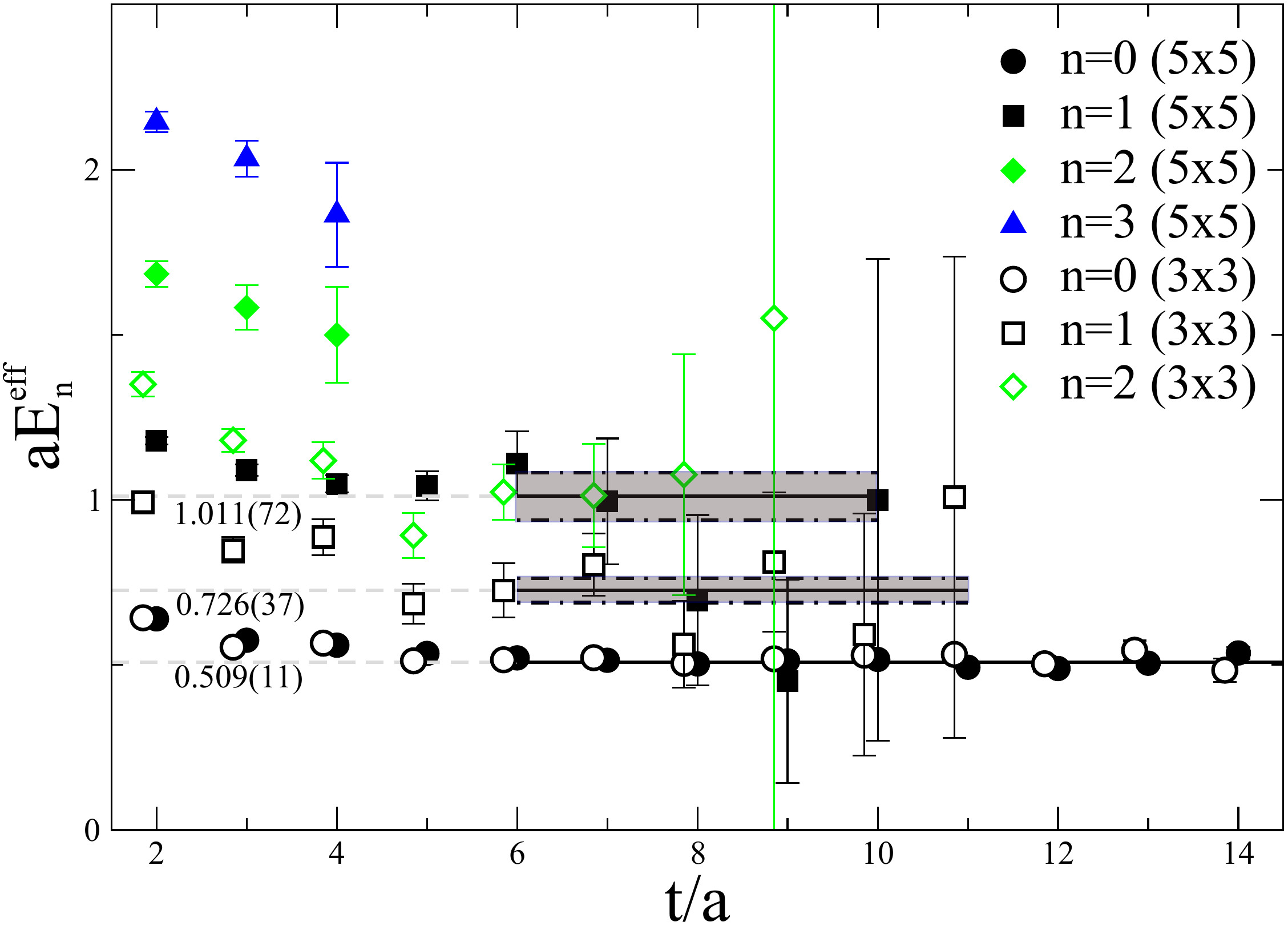}
  \caption{\label{spectrum} The spectrum when using $\chi_1^{(i)}$ at
    $\beta$=3.9, $a\mu$ = 0.004 (m$_\pi\sim308$~MeV) on a 32$^3$
    $\times$ 64 lattice.  For the 5$\times$5 GEVP we use 150
    configurations and {\it basis A}. For the $3\times$3 GEVP we use
    250 configurations with {\it basis B}.  The solid lines and bands
    show the fitted effective mass and jackknife error for the first
    excited state obtained from the two different GEVPs.}
\label{fig:5x5}
\end{figure}

From the above analysis it is clear that the merit of the variational
approach lies in the extraction of excited states, whereas the ground
state is equally well obtained using just a single smeared
interpolating function, in our case either $\chi_1^{(7)}$ or
$\chi_1^{(8)}$.  In Fig.~\ref{spectrum} we analyze the $5\times 5$
GEVP of {\it basis A} to extract the nucleon spectrum.  Despite the
low statistics used in this first examination we are able to obtain
effective mass plateaus $m_{\rm eff}(n)$ for the ground-state ($n=0$)
and the three excited states ($n=1$, $n=2$ and $n=3$), as has already
been done in other works~\cite{Mahbub:2009aa,Mahbub:2010rm}.
Fig.~\ref{spectrum} corroborates the previous observation that
including a heavily smeared interpolating field in the basis produces
an excited state with a lower energy.
Although increasing the level of smearing is essential for the
positive parity excited states, this is not the case when the negative
parity channel is considered, This issue will be discussed further in
the following section.

\subsection{Combining both $\chi_1$ and $\chi_2$}

In the preceding subsection we used a variational basis constructed
from different smearing levels of the $\chi_1$ interpolating field.
In this section, we extend the investigation by combining both
$\chi_1^{(i)}$ and $\chi_2^{(i)}$ each with two different smearing
levels
resulting in a $4\times 4$ correlation matrix.

\begin{figure}[h!]
  \centering \includegraphics[width=\linewidth,
    keepaspectratio]{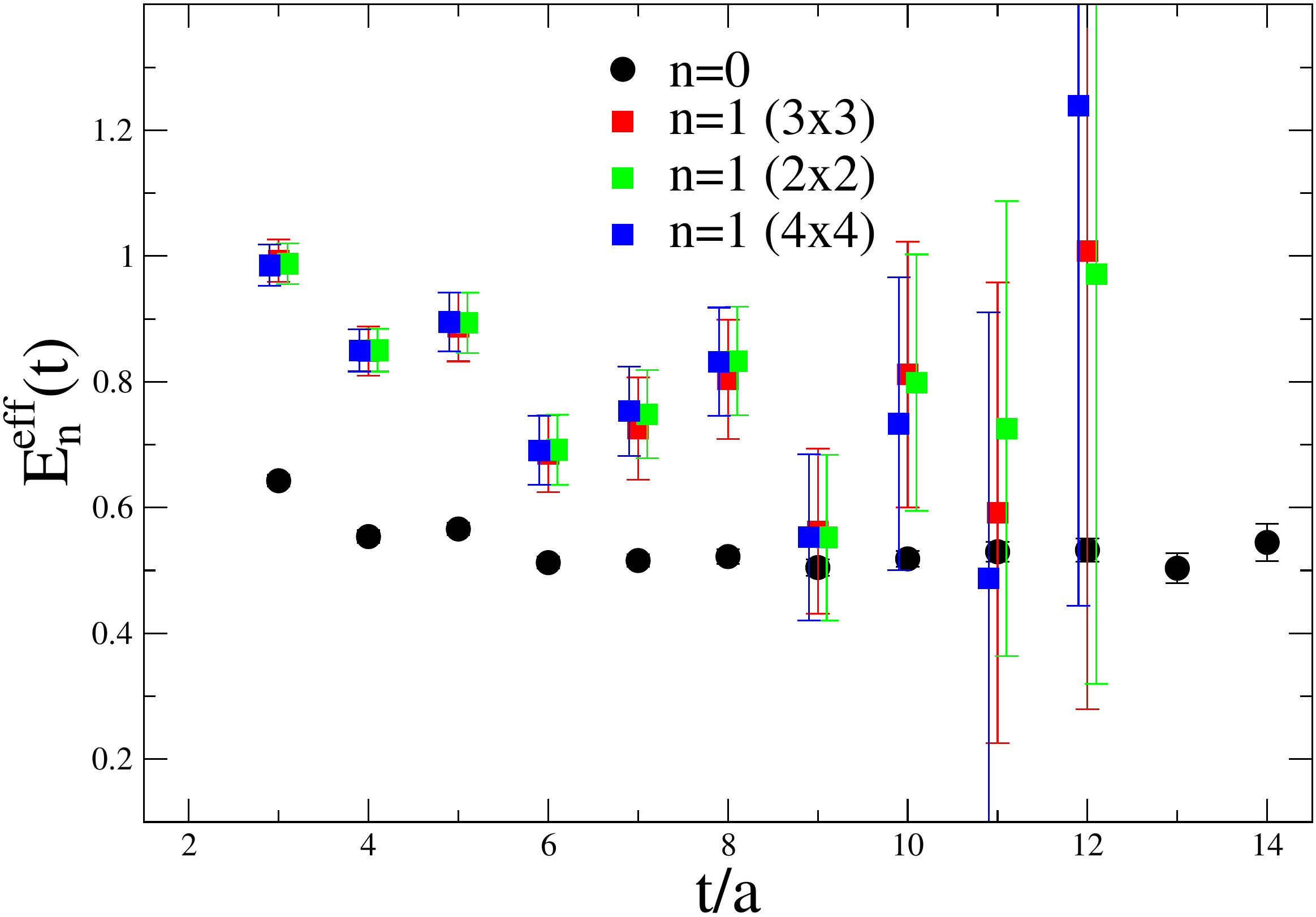}
  \caption{The effective mass for the ground and first excited states
    for the positive parity channel for $\beta$=3.9, a$\mu$ = 0.004 on
    a 32$^3$ $\times$ 64 lattice. The $3\times 3$ system is
    constructed using {\it basis B}; The $2\times 2$ system
    corresponds to $C_{1_i1_j}$ with $i,j=7,8$ and the $4\times 4$
    corresponds to $C_{a_ib_j}$ with $a,b=1,2$ and $i,j=7,8$.  250
    configurations are used. \label{Fig1_a}}
\end{figure}

For the positive parity channel we consider two different smearing
levels including the heavily smeared one that was found to give a
lower excited state energy, namely we consider $n_s$=50 and $n_s=180$
with $\alpha$=4.0 or correspondingly interpolating fields
$\chi^{(7)}_a$ and $\chi^{(8)}_a$ with $a=1,2$. In Fig.~\ref{Fig1_a}
we compare the results for the effective masses of the ground and
first excited state in the positive parity channel extracted using
this $4\times 4$ basis with those extracted from {\it basis B} of the
previous section (see Fig.~\ref{fig:5x5}). The effective mass plateaus
are statistically equivalent for both basis sets.





It is evident from the preceding analysis that the first excited state
can be obtained from the $2\times$2 GEVP using $C_{1_i1_j}$ with
$i,j=7,8$, or equivalently from the $4\times$4 GEVP using $C_{a_ib_j}$
with $a,b=1,2$ and $i,j=7,8$, a result that we will use in order to
further examine the first excited state for other ensembles. We note
that in both cases we use two different smearing levels.

Let us now examine the negative parity states. We first note that the
negative parity interpolating operator in Eq.~\ref{gevp1} has a
non-zero overlap with the two particle S-wave state that consists of a
nucleon and a pion. At the physical point, this state has lower energy
than the negative parity nucleon. To know \textit{a priori} at which
pion mass, the mass of the negative parity nucleon and the mass of the
$\pi\,N$ state cross requires knowledge of the pion mass dependence of
the negative parity nucleon.

\begin{figure}[h!]
\centering \includegraphics[width=\linewidth]{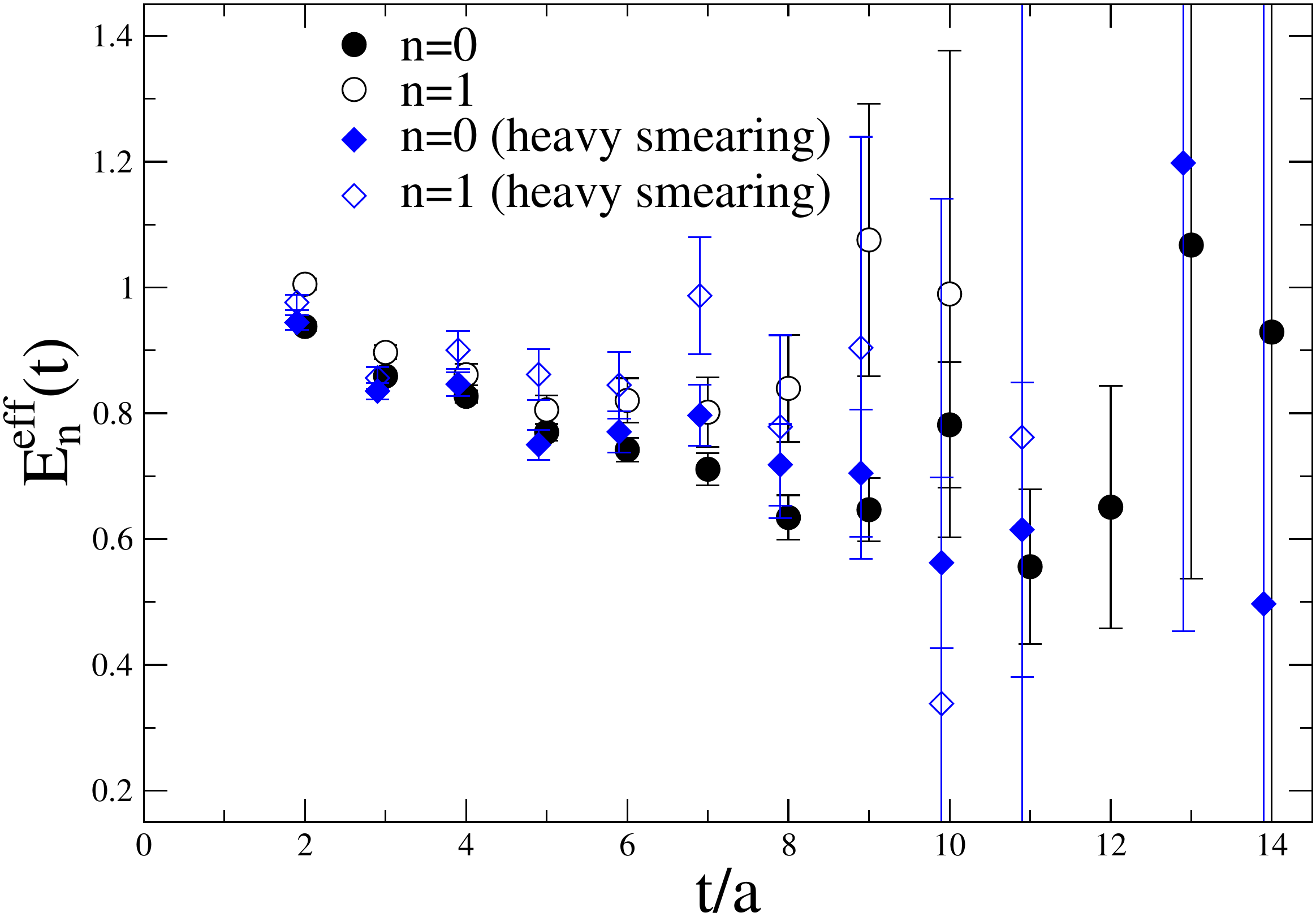}
\caption{The nucleon ground (filled symbols) and first excited states
  (open symbols) in the negative parity channel, evaluated via a
  $4\times4$ GEVP using two different basis sets:
  \{$\chi_1^{(1)}$,$\chi_1^{(5)}$,$\chi_2^{(1)}$,$\chi_2^{(5)}$\}
  (black circles) and the set
  \{$\chi_1^{(7)}$,$\chi_1^{(8)}$,$\chi_2^{(7)}$,$\chi_2^{(8)}$\}
  (blue diamonds).  250 configurations were used for this
  test.\label{neg_par_roleofsmearing}}
\end{figure}

\begin{figure}[h!]
\centering \includegraphics[width=\linewidth]{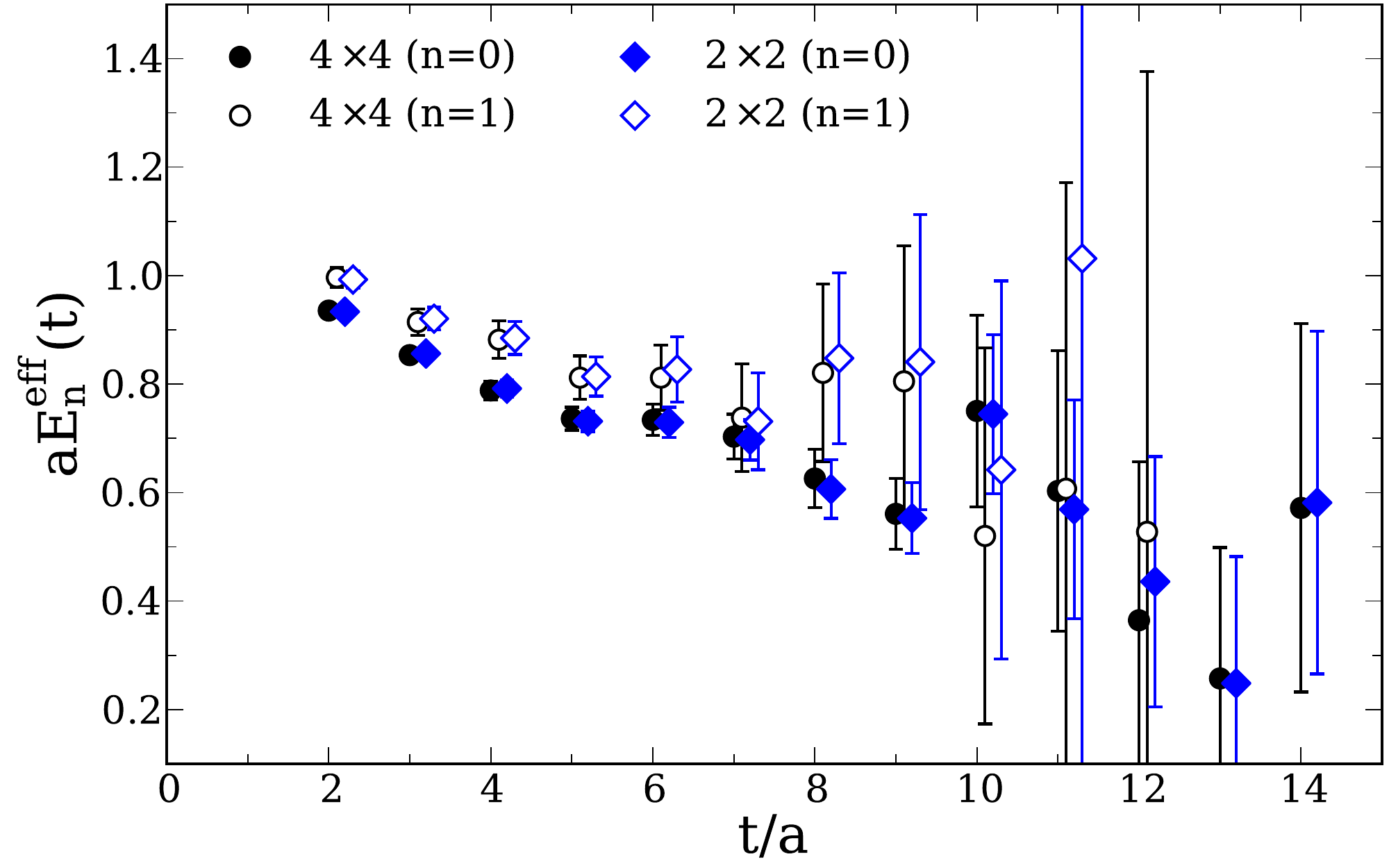}
\caption{The nucleon ground (filled symbols) and first excited states
  (open symbols) in the negative parity channel, evaluated using a
  $4\times4$ correlation matrix (black circles) and a $2\times2$
  correlation matrix (blue diamonds). The variational bases used are
  $\chi_a^{(i)}$, $a=1,2$ and $i=1,5$ and
  \{$\chi_1^{(5)}$,$\chi_2^{(5)}$\}.  250 configurations were used for
  this analysis.\label{fig:neg_par_4x4}}
\end{figure}

To explore the best variational basis for the negative parity channel
we carry out a similar analysis as with the positive parity channel.
We use two different bases each leading to a $4\times$4 correlation
matrix using both $\chi_1^{(i)}$ and $\chi_2^{(i)}$.  In the one set
we use $i=1,5$ while in the other $i=7,8$ i.e. the latter includes the
heavily smeared interpolating fields.  As is illustrated in
Fig.~\ref{neg_par_roleofsmearing}, including the heavily smeared
interpolator yields consistent results but with increased statistical
error.  In Fig.~\ref{fig:neg_par_4x4}, we show the ground and first
excited states obtained from a $4\times4$ and $2\times2$ GEVP. As in
the case of Fig.~\ref{neg_par_roleofsmearing}, the $4\times4$
correlation matrix is constructed using the basis $\chi_a^{(i)}$ with
$a=1,2$ and $i=1,5$, while the $2\times2$ using \{$\chi_1^{(5)},
\chi_2^{(5)}$\} (note that the basis \{$\chi_1^{(7)}, \chi_2^{(7)}$\}
yield equivalent results).  As can be seen, the two basis yield
results for the ground and first excited states that are statistically
equivalent.

Having verified that the $2\times2$ correlation matrix yields the same
energies for the ground and first excited states of the negative
parity as the $4\times4$ correlation matrix does, from here on, we
will use the $2\times 2$ basis to resolve the ground and first excited
negative parity states for all other pion masses. Knowing which one of
these is the multi-particle state would require investigation of the
dependence of the two energy levels on the lattice volume, which is
beyond the resources available to us for this work. Therefore, we
compare the two energy states with the sum of the nucleon and pion
mass, and from this infer which is the negative parity nucleon
state. Further examples of the effective masses extracted from the
$2\times 2$ correlation matrix are given in Figs.~\ref{L32T64_mu0.004}
and~\ref{fig:clov_meff}, discussed in the following section.

\section{The low-lying nucleon spectrum}
\label{sec:results}
In the previous section, we have shown that if we are interested in
the first excited positive parity states of the nucleon the
variational analysis using {\it basis B} is preferable to {\it basis
  A}.  Furthermore, we showed that the interpolating fields
$\chi_a^{(i)}$ with $a=1,2$ and $i=7,8$ suffice to determine the two
lowest state. Thus we construct a $4\times$4 correlation matrix, with
variational basis consisting of $\chi_1$ and $\chi_2$ with two
different smearing levels, one yielding a small rms radius and one a
large one.  The negative parity states were shown to be best extracted
from a $2\times$2 correlation matrix analysis, with a single level of
smearing using both interpolating operators (i.e. $\chi_1^{(7)}$ and
$\chi_2^{(7)}$).  We also note that results presented from here on
have been obtained with the statistics listed in
Table~\ref{Table:params}.

\begin{figure}[th!]
  \centering \includegraphics[width=\linewidth,
    keepaspectratio]{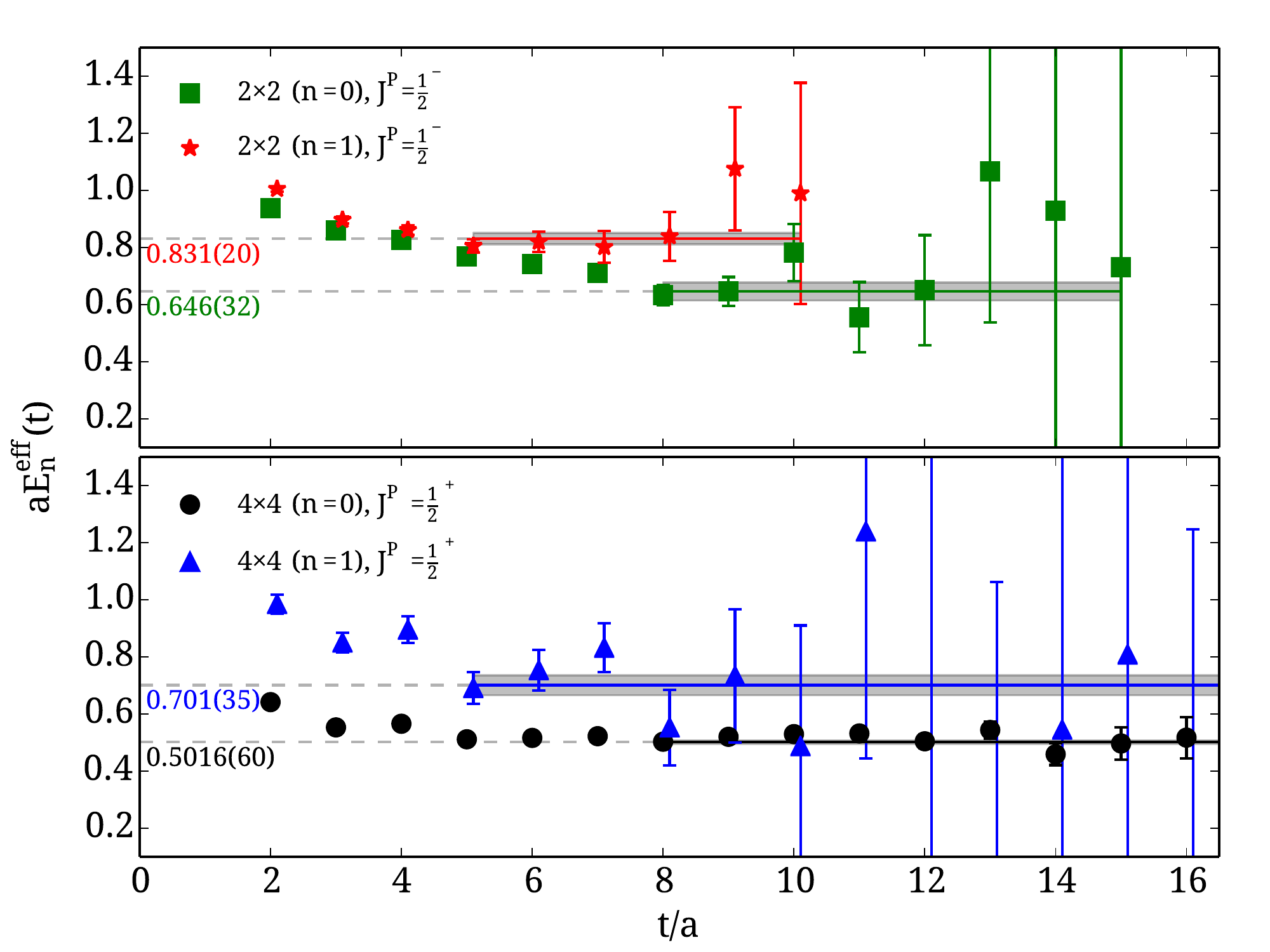}
  \caption{The effective masses of the two lowest lying nucleon states
    for the negative (upper panel) and positive (lower panel) states
    for the twisted mass ensemble with $\beta$=3.9, $a\mu$ = 0.004 and
    volume 32$^3$ $\times$ 64. For the positive parity states we use a
    $4\times 4$ correlation matrix with $\{\chi_1^{(7)}, \chi_1^{(8)},
    \chi_2^{(7)}, \chi_2^{(8)}\}$, while for the negative parity
    states we use a $2\times 2$ correlation matrix with $\chi_1^{(5)}$
    and $\chi_2^{(5)}$ as explained in the text.
    \label{L32T64_mu0.004}}
\end{figure}

\begin{figure}[th!]
  \centering \includegraphics[width=\linewidth,
    keepaspectratio]{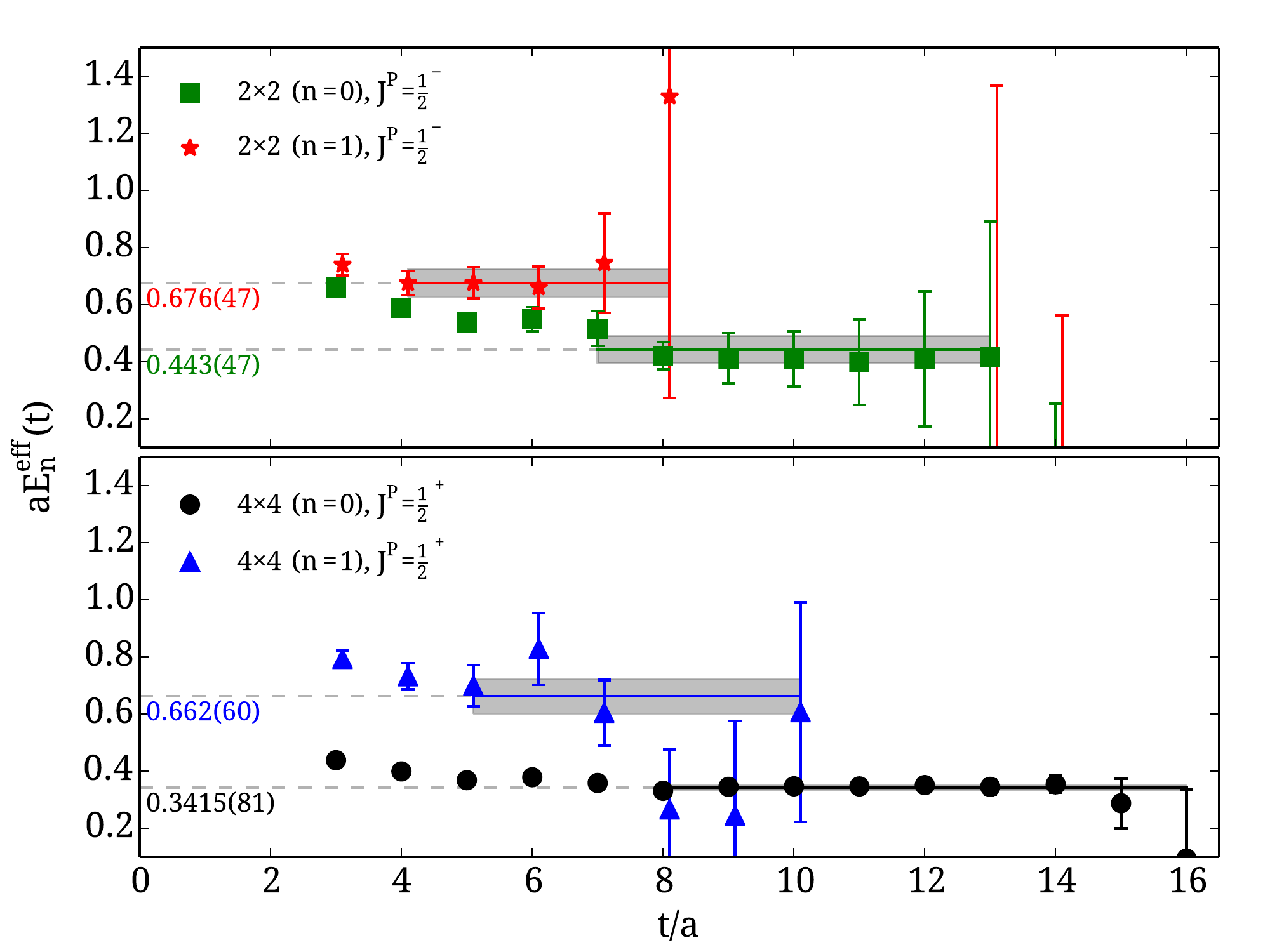}
  \caption{The effective masses of the two lowest lying nucleon states
    for the negative (upper panel) and positive (lower panel) states
    for the Clover ensemble. The notation is the same as in
    Fig.~\ref{L32T64_mu0.004} ~\label{fig:clov_meff}}
\end{figure}

In Figs.~\ref{L32T64_mu0.004} and~\ref{fig:clov_meff} we show the
effective masses for both positive and negative parity states, for a
twisted mass ensemble and for the Clover ensemble analyzed in this
work.  As can be seen, a plateau region can be identified for all
states.

\begin{figure}[h!]
  \includegraphics[width=\linewidth]{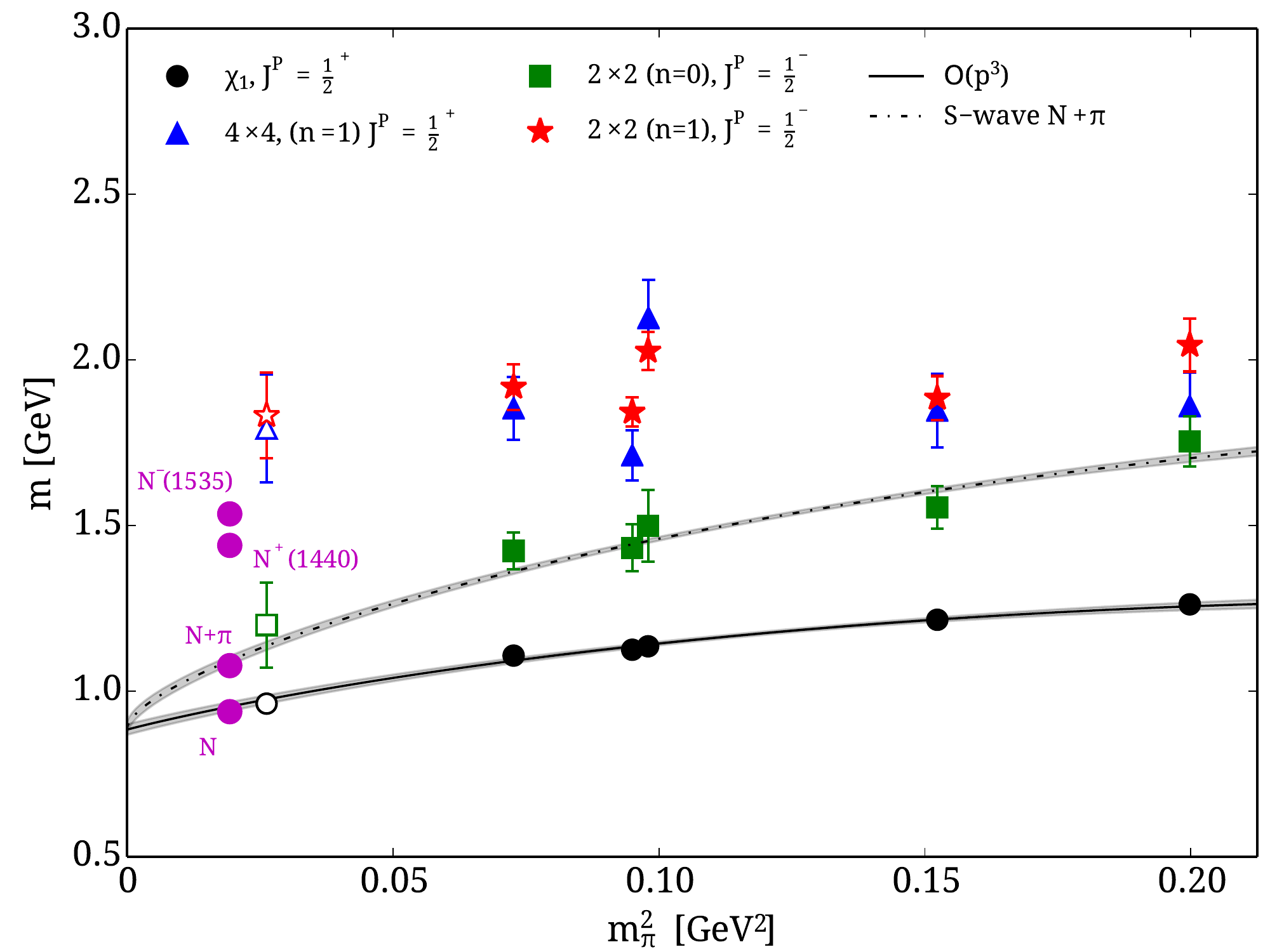}
  \caption{The first two positive and negative parity states measured
    on all gauge ensembles considered in this work. The twisted mass
    ensembles are plotted with filled symbols, while the results from
    the single Clover ensemble are denoted with the open symbols. We
    show chiral extrapolations for the nucleon ground state to ${\cal
      O}(p^3)$ as in Eq.~\ref{op3}, omitting the Clover point from the
    fit. The dashed line is a result of adding the pion mass to the
    ${\cal O}(p^3)$ curve. Physical masses for the different states
    are indicated by the magenta filled circles.
    \label{extrapolation}}
\end{figure}

The results for all of the ensembles of Table~\ref{Table:params} and
the single Clover ensemble are displayed in
Fig.~\ref{extrapolation}. For the nucleon mass we apply continuum
chiral perturbation theory to extrapolate lattice results to the
physical pion mass, omitting the Clover point from the fit.  We use
SU(2) heavy baryon chiral perturbation theory to ${\cal O}(p^3)$ given
by \be m _N(m_\pi) = m^{(0)}_N-4c^{(1)}_N \;m_{\pi}^2 - \frac{3 g_A^2
}{16\pi f_\pi^2} \;m_\pi^3.
\label{op3}
\ee Since the lattice spacing was fixed using the nucleon mass for the
twisted mass ensembles it is no surprise that the curve passes through
the physical value.  Since the Clover point was not included in the
fit the fact that it lies on the curve provides a consistency check
for our procedure.  In the figure we also show a curve obtained by
adding the pion mass to the nucleon mass. As can be seen, for all pion
masses considered here, the negative parity ground state is consistent
with the mass of the pion plus nucleon, indicating that this is the
two particle $\pi$~N state in an S-wave configuration. We also observe
that the first excited states in the positive and negative channels
remain close together for all pion masses.
 
\begin{figure}[h!]
  \centering \includegraphics[width=\linewidth,
    keepaspectratio]{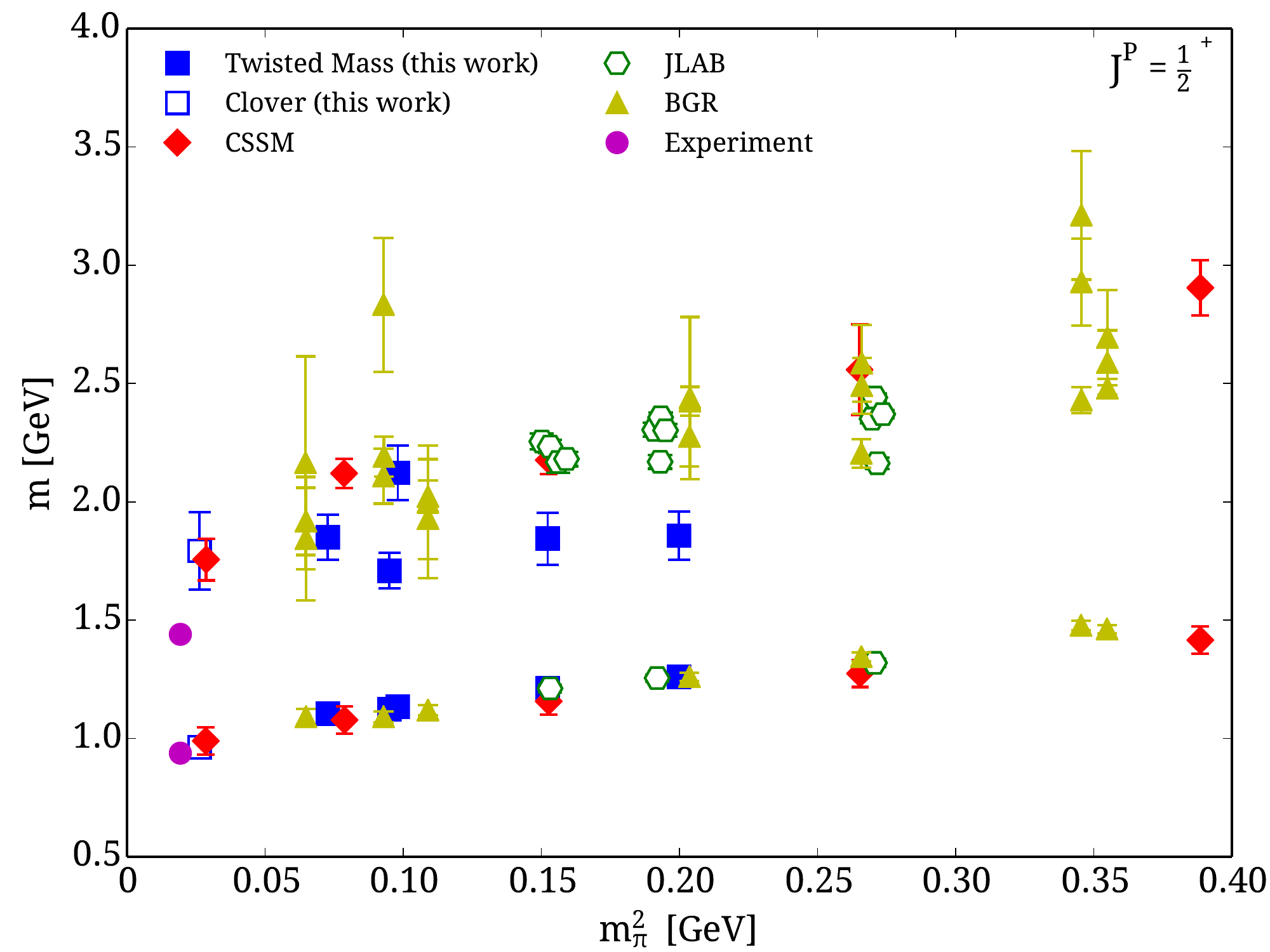}
  \caption{\label{compare_pos} The positive parity states of this work
    (filled and open squares) compared with results from other groups,
    that include a $N_f=2+1$ Clover improved fermion calculation by
    the CSSM collaboration~\cite{Mahbub:2012zz} (red diamonds), a
    calculation using anisotropic clover lattices by the Hadron
    Spectrum Collaboration~\cite{Edwards:2011jj} (open hexagons) and a
    calculation using the Chirally Improved Dirac Operator by the
    Bern-Graz-Regensburg (BGR) collaboration~\cite{Engel:2013ig}
    (yellow triangles).  }
\end{figure}

\begin{figure}[h!]
  \centering \includegraphics[width=\linewidth,
    keepaspectratio]{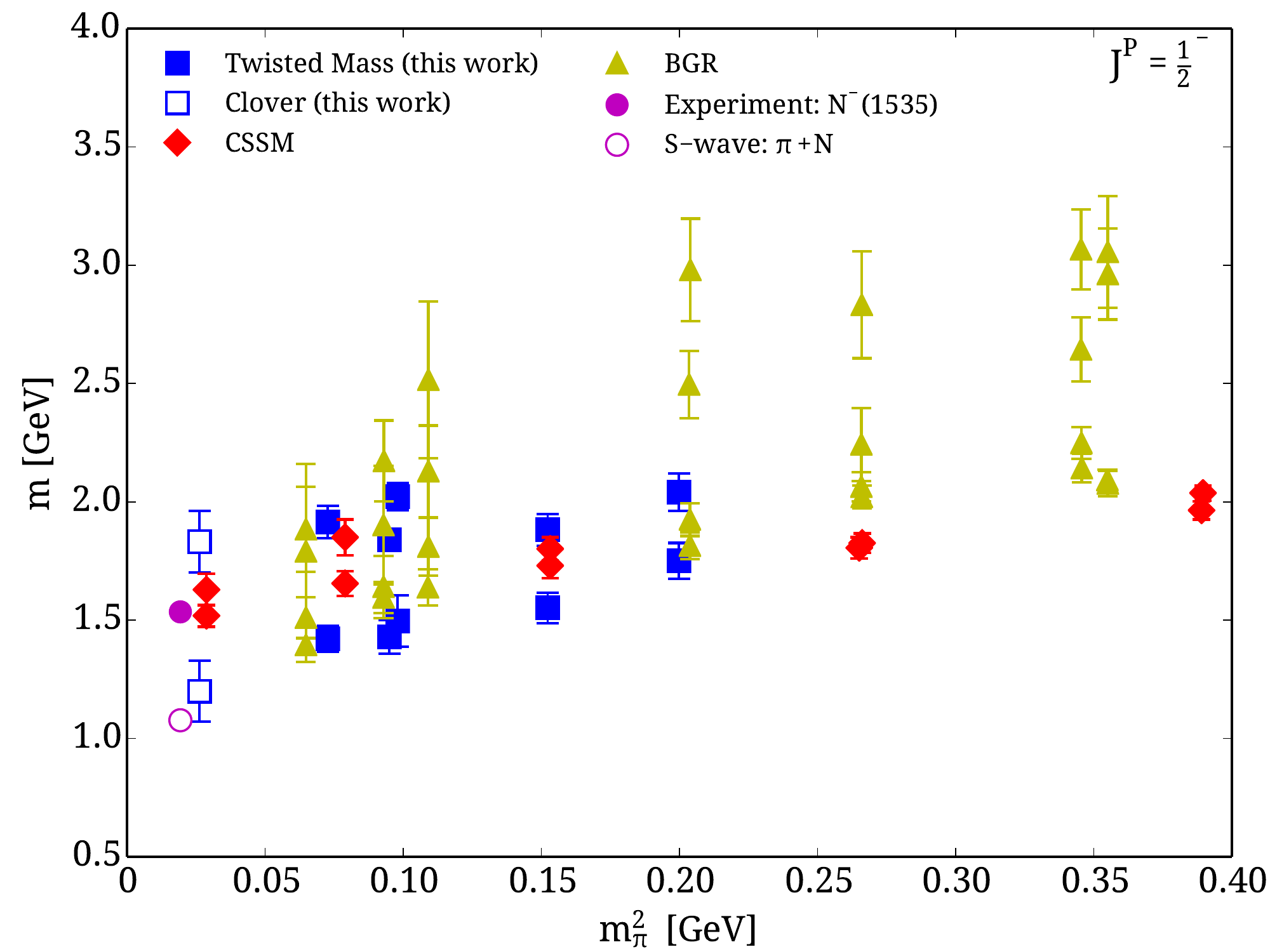}
  \caption{\label{compare_neg} The negative parity states of this work
    compared with calculations from other groups. The CSSM results are
    from~\cite{Mahbub:2013ala}, while the rest of the notation is as
    in Fig.~\ref{compare_pos}.}
\end{figure}

In Figs.~\ref{compare_pos} and~\ref{compare_neg} we compare the
results of this work with three other calculations available in the
literature. Namely, we compare with the results obtained using a
Clover improved fermion action by the CSSM
collaboration~\cite{Mahbub:2012zz} with $a\simeq0.09$~fm, a
calculation using anisotropic Clover lattices by the Hadron Spectrum
Collaboration~\cite{Edwards:2011jj} with spatial lattice spacing
$a_s=0.123$~fm and a calculation using the Chirally Improved Dirac
Operator by the BGR collaboration~\cite{Engel:2013ig} and lattice
spacings between $0.13$ and $0.14$~fm.  We note that the lattice
spacings for the two latter calculations are notably larger than those
used in this work arising issues about cut-off effects.

The first observation is that all lattice results are in reasonable
agreement for the ground state energies of both parity channels. The
second major observation is that our data for the first excited state
of the nucleon in the positive parity channel, although consistent at
near physical pion mass with the other lattice calculation at similar
pion mass, namely that from the CSSM Collaboration, is still higher
than the experimentally measured mass for the Roper. Given that our
lattice volume is comparable to that of Ref.~\cite{Mahbub:2012zz}
volume effects can be responsible for the larger values. In the
negative parity channel our results are consistent with the ones from
the BGR collaboration. We can clearly see that for all pion masses
considered the negative parity ground state is consistent with a
$\pi\,N$ state in an S-wave. To the statistical accuracy available to
us, the first excited negative parity state appears to be converging
to $N^-(1535)$, however the errors are too large to draw concrete
conclusions. Overall, the early loss of signals seen in the plateaus
of the excited states shown in Figs.~\ref{L32T64_mu0.004}
and~\ref{fig:clov_meff} indicates that a high statistics calculation
of these quantities is merited using e.g. recently developed
noise reduction techniques~\cite{Blum:2012uh}.

\section{Conclusions}
\label{sec:conclusions}
In this work we apply the variational method to investigate the
excited states of the nucleon. Two sets of variational bases are used
and the analysis of the resulting GEVP was performed using the
standard approach of fixing $t_0$ as well as by varying $t_0$ such
that $t_0\ge t/2$ as advocated in Ref.~\cite{Blossier:2009kd}. Within
the current statistical accuracy, we found that for the nucleon
excited states no observable improvement is obtained as compared to
fixing $t_0$. Limiting ourselves to the first excited state of the
nucleon in the positive parity channel requires a combination of one
broadly and one narrowly smeared interpolating field. Including both
$\chi_1$ and $\chi_2$ yields a $4\times 4$ correlation matrix, which
we use to extract results in the positive parity channel for a number
of $N_f=2$ twisted mass fermion ensembles.  Besides the twisted mass
fermion ensembles we use in addition an $N_f=2$ clover fermion
ensemble with pion mass almost equal to the physical value. At this
lightest pion mass of 160~MeV we find an excited state, which is still
higher than the Roper but consistent with another calculation at
similar pion mass from the CSSM collaboration. We do not observe a
strong pion mass dependence and the higher value may be due to finite
volume effects, which must be further investigated. In the negative
parity channel we obtain results that reveal the $\pi N$ scattering
state and an excited state, which at $m_\pi=160$~MeV is still higher
than the physical value of $N^-$. It is clear from this analysis that
extracting the excited states is still a challenge and more work is
needed to understand the low-lying spectrum of the nucleon.

\section*{Acknowledgments}

For the numerical calculations, we have used the Cy-Tera facility of
the Cyprus Institute under the Cy-Tera first access call (Project No.
lspro113s1) and third access call (Project No. lsprob115s1). The
Cy-Tera project is funded by the Cyprus Research Promotion Foundation
under Contract No. NEA Y$\Pi$O$\Delta$OMH/$\Sigma$TPATH/0308/31. In
addition, this work was granted access to the HPC system ``Lindgren'' of
KTH Stockholm made available within the Distributed European Computing
Initiative by the PRACE-2IP receiving funding from the European
Community's Seventh Framework Programme (FP7/2007-2013) under Grant
No. RI-283493. We thank the staff members of these sites for their
kind and sustained support. This work is supported in part by the
Cyprus Research Promotion Foundation under Contract No. KY-Γ/0310/02/,
the Research Executive Agency of the European Union under Grant
No. PITN-GA-2009-238353 (ITN STRONGnet) and the FP7
infrastructures-2011-1 project HadronPhysics3 under Grant No. 283286.





\bibliography{ref}







\end{document}